\documentclass[sigconf]{acmart}

\usepackage{tabu}
\usepackage{float}
\usepackage{gensymb}
\usepackage{textcomp}
\usepackage{amsmath}
\usepackage{algpseudocode}
\usepackage{pifont}
\usepackage{verbatim}
\usepackage{graphicx}
\usepackage{hyperref}
\usepackage{seqsplit}
\usepackage{caption}
\usepackage{enumitem}
\usepackage{url}
\usepackage{booktabs}
\usepackage{mathtools}
\usepackage{multirow}
\usepackage{tabularx}
\usepackage{bigstrut }
\usepackage{longtable}
\usepackage{ltxtable}
\usepackage{xtab,afterpage}
\usepackage{lipsum}
\usepackage[ruled]{algorithm2e}
\usepackage{subcaption}
\usepackage{pbox}
\usepackage{framed}
\usepackage{array}  
\usepackage[capitalise,noabbrev]{cleveref}
\usepackage{adjustbox}
\usepackage{balance}

\newcommand{\checkmarkR}{\ding{51}}%
\newcommand{\xmark}{\ding{55}}%

\copyrightyear{2020} 
\acmYear{2020} 
\setcopyright{acmcopyright}
\acmConference[AutoSec'20]{The 2nd ACM Workshop on Automotive and Aerial Vehicle Security}{March 18 2020}{New Orleans, LA, USA}

\begin{document}

\title{Security and Privacy Challenges in Upcoming Intelligent Urban Micromobility Transportation Systems}

\author{Nisha Vinayaga-Sureshkanth}
\affiliation{
\department{Department of Computer Science}
\institution{University of Texas at San Antonio}
\streetaddress{1 UTSA Circle}
\city{San Antonio}
\state{Texas}
\postcode{78249}
\country{USA}
}
\email{vsnisha@ieee.org}

\author{Raveen Wijewickrama}
\affiliation{
\department{Department of Computer Science}
\institution{University of Texas at San Antonio}
\streetaddress{1 UTSA Circle}
\city{San Antonio}
\state{Texas}
\postcode{78249}
\country{USA}
}
\email{raveen.wijewickrama@utsa.edu}

\author{Anindya Maiti}
\affiliation{
\department{Institute for Cyber Security}
\institution{University of Texas at San Antonio}
\streetaddress{1 UTSA Circle}
\city{San Antonio}
\state{Texas}
\postcode{78249}
\country{USA}
}
\email{a.maiti@ieee.org}

\author{Murtuza Jadliwala}
\affiliation{
\department{Department of Computer Science}
\institution{University of Texas at San Antonio}
\streetaddress{1 UTSA Circle}
\city{San Antonio}
\state{Texas}
\postcode{78249}
\country{USA}
}
\email{murtuza.jadliwala@utsa.edu}

\renewcommand{\shortauthors}{Vinayaga-Sureshkanth et al.}

\begin{abstract}
Micromobility vehicles are gaining popularity due to their portable nature, and their ability to serve short distance urban commutes better than traditional modes of transportation. Most of these vehicles, offered by various micromobility service providers around the world, are shareable and can be rented (by-the-minute) by riders, thus eliminating the need of owning and maintaining a personal vehicle. However, the existing micromobility ecosystem comprising of vehicles, service providers, and their users, can be exploited as an attack surface by malicious entities -- to compromise its security, safety and privacy. In this short position paper, we outline potential privacy and security challenges related to a very popular urban micromobility platform, specifically, dockless battery-powered e-scooters.  
\end{abstract}

%why are they popular?
%why are they a threat?
%what are you doing?

\begin{sloppypar}
\begin{CCSXML}
<ccs2012>
<concept>
<concept_id>10003120.10003138.10003141</concept_id>
<concept_desc>Human-centered computing~Ubiquitous and mobile devices</concept_desc>
<concept_significance>300</concept_significance>
</concept>
<concept>
<concept_id>10010520.10010553</concept_id>
<concept_desc>Computer systems organization~Embedded and cyber-physical systems</concept_desc>
<concept_significance>100</concept_significance>
</concept>
</ccs2012>
\end{CCSXML}

\ccsdesc[300]{Human-centered computing~Ubiquitous and mobile devices}
\ccsdesc[100]{Computer systems organization~Embedded and cyber-physical systems}

\keywords{Micromobility, e-scooters, security, privacy}

\maketitle
\end{sloppypar}

\section{Introduction}
\label{intro}

Micromobility has emerged as a popular mode of urban transportation, and collectively represents the compact, lightweight vehicles such as electric scooters, electric skateboards, electric bikes, hoverboards, segways, etc. \cite{nacto}. Among these vehicles, electric scooters or e-scooters are the most appealing to urban users \cite{sae}, mainly due to the shared or rent-by-the-minute schemes offered by a number of different service providers. These vehicles are preferred by users for a plethora of reasons, such as their portable nature which allows easier bypassing of urban traffic, and their ability to reach destinations that otherwise required walking \cite{shed}. While the docked models can be parked at fixed locations, the dockless models can be dropped off at a more flexible location. The most noteworthy aspect is their potential to connect the gray area between traditional points-of-interests such as parking lots and bus stops, and final destinations such as workplaces and campus buildings, in congested areas or places with limited transportation. These convenient options also save users from maintenance costs associated with owning a vehicle, and from rather expensive ride-hailing or ride-sharing costs for short distance travels. Further, the streamlined process of geo-locating nearby scooters through the service provider's application, easy payment options, and flexible drop-off or parking options make micromobility e-scooter services notably attractive to urban commuters.

The electric scooter adoption has either been a success (coexisting with existing modes) or a failure (creating chaos) in urban communities depending on the readiness of the cities to these unconventional transportation means \cite{dockless94-online,micromob51-online,bikeandS52-online,sharedUs28-online}. Shared electric scooters deployed by the service providers, such as Lime and Bird, are almost universally equipped with an embedded controller and can be activated using their corresponding smartphone application \cite{bird2019, lime2019, lyft2019, jump2019}. The scooters communicate with the smartphone application using BLE (Bluetooth Low Energy) spectrum and/or using the Internet, both of which are well established and prevalent technologies. However, use of such communication channels also opens the door to a plethora of attacks, some of which can be especially effective on micromobility e-scooters. Similarly, the use of cloud for managing the e-scooter rental and user data can become a lucrative target. The literature is already rich with several different attacks on micromobility e-scooters, and some of the attack resources are even readily accessible online. Amateurs without expert knowledge or grasp of the underlying technology can easily adopt and execute (with reasonable success) some of these attacks for monetary or other gains, with only slight modifications of existing tools and techniques. 

While some of the attacks have already been addressed, many remain unaddressed either due to lack of a suitable solution or due to lack of awareness. Moreover, while many different attacks are already published, no prior work systematized the various attack points within the micromobility ecosystem and the different types of adversaries. In this paper, we systematize and discuss security and privacy concerns pertaining to the micromobility e-scooter services, and attack scenarios plausible with their interfaces. Such a systematize discussion will help developers and researchers to easily identify weaknesses, and improve overall security and privacy properties of the ecosystem. We also discuss potential countermeasures against some of the attacks, wherever possible. Before describing the different attacks in \cref{obs}, we first detail the micromobility e-scooter service ecosystem in \cref{bgr}.

\section{Background}
\label{bgr}
Micromobility e-scooter users can access the shared vehicles via the smartphone application provided by the service providers. First, the rider creates a user account and register a payment method with the service provider. The rider can then skim through a list of nearby e-scooters through the application, and navigate to the desired e-scooter. Once in close proximity, the rider scans the QR code on the e-scooter, initializing the riding process and starting the e-scooter both contingent on the funds (credits) and charge remaining on the e-scooter battery. The service providers charge from 15 to 30 cents per minute to ride the e-scooter along with a base activation fee. Certain e-scooter models can travel up to a distance of 28 $miles$ in a single charge, which is much shorter than the average distance covered in e-scooter trips \cite{hardt2019usage}. The e-scooters can cover up to 18.6 \textit{miles per hour} in a typical ride based on the road conditions and other vehicle traffic, and are equipped with headlights, tail or brake lights, a bell or horn and sometimes a display. While some models have throttle and front brakes located on the handle bar similar to a motorcycle, other models have foot controlled brakes or rear disk brakes or anti-lock brakes for a safer ride \cite{razor2019, spin2019,bird2019, lime2019, lyft2019, jump2019}. %In addition certain advanced models also has a display which provides basic e-scooter/ride related information \cite{}. 

\begin{figure}[t]
\centering
\includegraphics[width=0.8\linewidth]{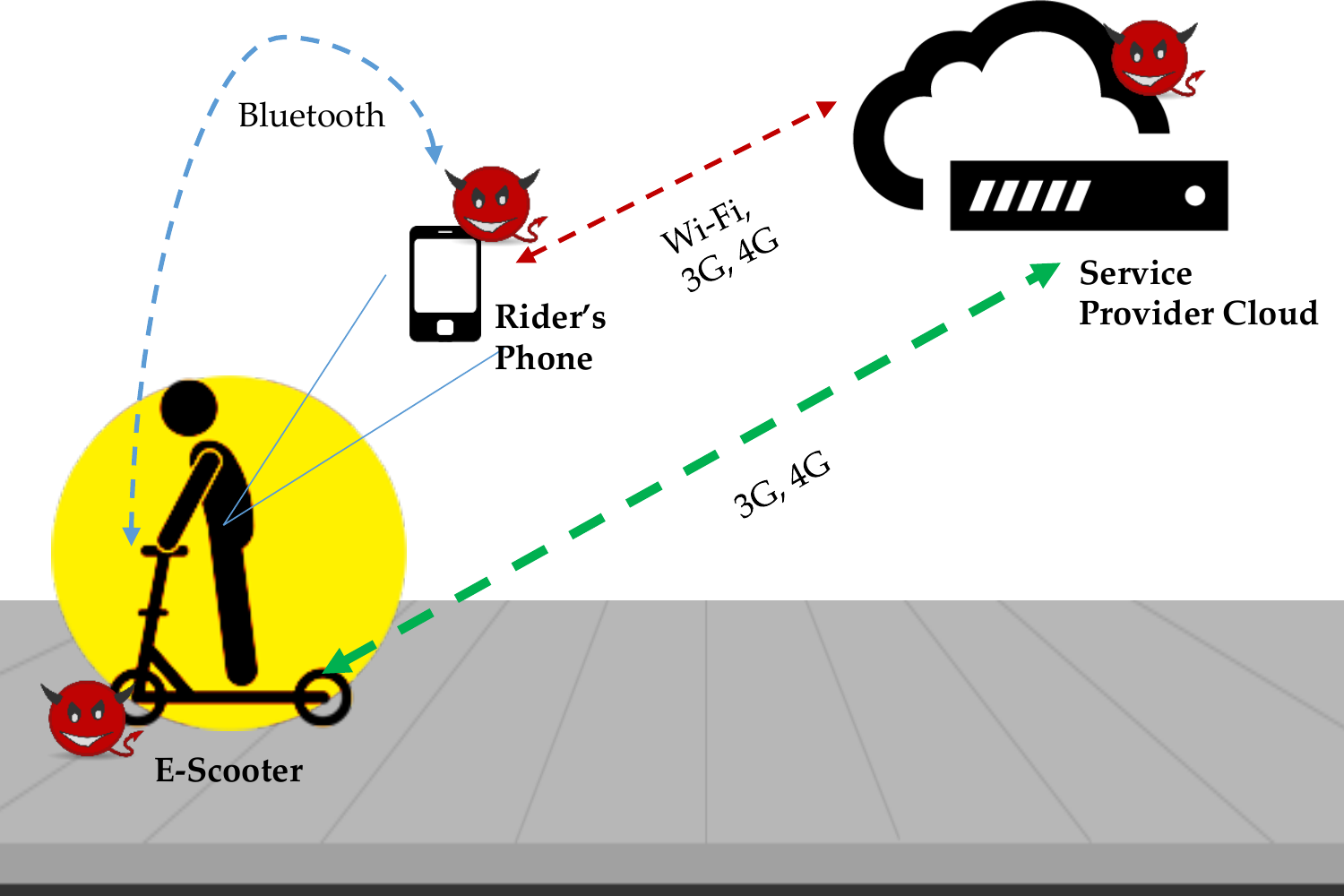}
\captionof{figure}{E-scooter ecosystem and attack points.}
\label{fig:ble-example}
\centering
\vspace{-0.1in}
\end{figure}

E-scooters primarily rely on BLE, which broadcasts packets at fixed intervals that can be captured by smartphones. These packets contain unique identifiers which helps in identifying the e-scooters from other BLE devices. To start or stop most e-scooters, the operating smartphone needs to have both Bluetooth enabled and Internet data available. This can be easily tested by toggling both the options on and off and checking if the e-scooter can be started or stopped with only one of the features. Overall, the important entities that are part of the micromobility infrastructure are the rider, the smartphone used by the rider to communicate with the e-scooter, the offline (BLE) and online (cloud) communication medium used by the service providers, and the e-scooter itself. The attacker or adversary, active or passive, in a micromobility ecosystem can be a \textit{rider}, an \textit{outsider} or the \textit{service provider} as shown in \cref{fig:ble-example}.

A \textbf{rider} can manipulate the e-scooter and the service provider for personal benefits by exploiting vulnerabilities in the smartphone application and the communication channels.
An \textbf{outsider} can be any third party entity with harmful (or deceptive) intentions towards the rider or the service provider or  an entity who may be curious about the micromobility ecosystem or the rider. The outsider can attack the e-scooter, the rider, the communication channel or the service provider.
The \textbf{service provider} can obtain information about the micromobility users and their surroundings in addition to e-scooter operations. As a result, the service provider may constantly monitor user habits and preferences and share sensitive information with third parties to maximize revenue.
These entities cause various hazardous scenarios and privacy concerns that affect the micromobility e-scooter ecosystem intentionally or unintentionally. A comprehensive set of security and privacy concerns are systematically described next.

\section{Potential Attacks}
\label{obs}

%Many tools are available to populace to execute these attacks: SDR, USB (Facedancer), JTAG (BusPirate), Oscilloscope, logic analyzer (Salae).

\subsection{Physical Damage} 
\subsubsection{Observations} 
The key components in the e-scooters are its battery, engine, brakes, headlight, controller chip and other mechanical, electrical and electronic components to ensure safe and smooth driving experience for the rider. Some e-scooters also have anti-theft mechanisms in place such as physical locks that can be enabled and disabled from the rider's smartphone through BLE. They may also have alarm systems that emit loud sounds or frequent beeps in the event they are displaced without unlocking them. Any of these components can be the target of an attacker (a rider or an outsider) and often times the e-scooter itself. For instance, e-scooter brake wires and batteries were targeted in several physical attacks \cite{Floridam83:online}. E-scooters were also stolen and possibly used as a personal e-scooter by flashing custom firmware on the e-scooter controller \cite{customFirmware:online,caliScooterTheft:online,scooterTheftSol:online}.

\subsubsection{Consequences} 
The attacker (the rider or outsider) can target the e-scooter battery, specifically drain it before attempting to move or acquire it, in order to circumvent the security mechanisms. Once the e-scooter is acquired, the attacker can install malicious modules, remove or replace key components before placing it back in the streets to control the e-scooter remotely or to covertly gather data about the e-scooter and populace near the e-scooter. These tampered e-scooters can be a threat to road users in many ways. The attacker can intentionally injure the victim rider by remotely manipulating or interfering on the with the e-scooter's brakes, damaging the tires or other physical damage that could incapacitate the e-scooter. The attacker can indirectly target a group of non-riders or other vehicles on road by strategically targeting a rider in their path. Such attacks cause monetary loss for the service providers, environmental pollution (when burned by vandalists), and physical damage to road users. The attacker, with financial interests, can sell the untraceable (removing key identifiers) or modified (malware) e-scooters online, which can later be converted for personal use \cite{customFirmware:online}.

\subsubsection{Countermeasures} 
The service providers could assign the e-scooter chargers (users tasked with charging the e-scooters for a small payment) an additional task of checking the vital functionalities of the e-scooters such as brakes and tires before redeploying them after charging. Further, the non-riders and other road users who come across any issues in the e-scooter could have an option to report them allowing the service provider to take it out of service in order, thus preventing any other user from riding it.

\subsection{Eavesdropping} 
\label{api}
\subsubsection{Observations}
The e-scooters communicate with the rider smartphone over the BLE channel, and in some cases over the Internet. Entities can listen to data exchanges between the e-scooter and rider smartphone over these channels with suitable hardware (Ubertooth) or software (Wireshark). 
\subsubsection{Consequences} 
The ability to successfully sniff control commands and service requests can give leverage to the attacker, active or passive, to study the micromobility e-scooter ecosystem when combined with fuzzing (\cref{fuzz}) and to explore a world of exploitations described in \cref{mitm,dos}. For instance, researchers in \cite{AppAnaly} identified Bird's API endpoints that contained the QR code information used to reserve e-scooters and make them chirp without being in physical proximity to the e-scooter. Also, the leakage of the e-scooter lock information from the API shows a possibility of attackers stealing them without draining e-scooter batteries.

\subsubsection{Countermeasures}
Though it is difficult to deter attackers from sniffing data, the service provider could design the applications and services to prevent sensitive information leakages by disabling or making such features inaccessible to unauthorized entities.

\begin{table*}[]
\caption{User data collected, according to the privacy policies of various micromobility service providers.}
\begin{adjustbox}{width=1\textwidth}
% \begin{tabular}{@{}|l|l|l|l|l|l|l|l|l|@{}}
\label{tab:policy}
\begin{tabular}{@{}|c|c|c|c|c|c|c|c|c|@{}}
\toprule
Service Provider & \multicolumn{4}{c|}{User Provided}                                   & \multicolumn{3}{c|}{Automatically Collected}        & From 3rd Party Sources                            \\ \midrule
                 & Contact Info & Billing Info & Identification Info & Demographic Info & Device Info & Location and Vehicle Info & Analytics & User interactions related to the service provider \\ \midrule
Bird             &     \checkmarkR         &      \checkmarkR        &        \checkmarkR             &       \checkmarkR           &     \checkmarkR        &         \checkmarkR                  &     \checkmarkR      &       \checkmarkR                                            \\ \midrule
Lime             &     \checkmarkR         &      \checkmarkR        &        \checkmarkR             &       \checkmarkR           &    \checkmarkR         &       \checkmarkR                    &    \checkmarkR       &      \checkmarkR                                             \\ \midrule
Razor            &     \checkmarkR         &     \xmark         &      \checkmarkR               &     \xmark             &      \checkmarkR       &          \checkmarkR                 &     \checkmarkR      &     \checkmarkR                                              \\ \midrule
Lyft             &     \checkmarkR         &       \checkmarkR       &       \checkmarkR              &      \checkmarkR            &    \checkmarkR         &      \checkmarkR                     &    \checkmarkR       &     \checkmarkR                                              \\ \midrule
Jump             &     \checkmarkR         &     \checkmarkR         &         \checkmarkR            &      \checkmarkR            &     \checkmarkR        &        \checkmarkR                   &   \checkmarkR        &      \checkmarkR                                             \\ \bottomrule
\end{tabular}
\end{adjustbox}
\end{table*}

\subsection{Man-in-the-Middle and Replay Attacks} 
\label{mitm}
\subsubsection{Observations}
With sufficient knowledge obtained from the eavesdropping attack, the attacker can intervene (modify commands or drop data) communication between a rider smartphone and an e-scooter. BLE vulnerabilities have allowed researchers to perform MITM attacks on the Xiaomi M365 e-scooter \cite{cameron2019iot,speedHack:online,msgHack:online}. 
\subsubsection{Consequences} 
The ability to successfully replay or mimic a rider opens up the possibilities of attacks described in \cref{dos,fuzz}. The attacker can jam the communication medium by bombarding with multiple legit or malformed requests (replay) in a short time period for a Denial of Service or Fuzzing attack, ultimately exhausting e-scooter battery. The attacker (with malicious intent towards the rider) can control the e-scooter by injecting commands remotely, and intentionally cause physical harm to the rider and other road users alike. 
\subsubsection{Countermeasures} 
The attacks can be mitigated by observing the the behavior of the nodes and the nature of their packet requests and blacklisting suspicious nodes \cite{guo2017efficient,conti2016survey}. Furthermore, the service providers and micromobility users should consider keeping their services and smartphones updated to patch up any security vulnerabilities.

\subsection{Denial-of-Service} 
\label{dos}
\subsubsection{Observations}
This type of attacks has the ability to disrupt any service such as locking and unlocking the e-scooter, etc. (often rendering them inaccessible) and can be targeted towards the e-scooter exhausting its resources, or towards the service providers affecting their quality of service.
\subsubsection{Consequences} 
Combined with the findings from \cref{api}, the attacker can implement the attack in a massive scale, plausibly paving way for a DDoS attack. For instance, the attacker could target a particular service provider and intentionally cause monetary loss for the service provider by preventing riders from using their services in a targeted area through a worm hole or black hole infiltration of modified e-scooters. 
\subsubsection{Countermeasures} 
The service provider could systematically monitor and filter real time traffic, maintain logs and implement disaster recovery plans for quick recovery in the event of an attack \cite{zhang2016denial,thai2016resiliency}.

\subsection{Spoofing} 
\label{scoot}
\subsubsection{Observations}
The micromobility applications track the location of the e-scooter using the inbuilt GNSS module on board the e-scooter or using the riders smartphone or both. An attacker, the rider or the outsider, can target either options to manipulate this location. In the first approach, the rider can install any location spoofing applications (available on the Internet) on the smartphone to fake their location \cite{fakeGPSLocLW-online, bestFakeGPSApps-online, zhao2017location}. After installation, the rider can easily trick the micromobility application and the service provider. In the second approach, the attacker can manipulate or replay GPS signal using SDR hardware (HackRF, USRP, BladeRF, etc.) which can produce and broadcast forged GPS signals to the victim receiver. It is also possible for an attacker to capture a GPS signal from a different location and rebroadcast it to the victim receiver (replay attack) \cite{zeng2014location}. The latter approaches can trick both the GNSS modules on the smartphone (solely reliant on GPS for location) and the e-scooter. 

\subsubsection{Consequences} 
A successful location attack gives the attacker the ability to manipulate the application and navigate the victim who depends on GPS navigation into dangerous situations or locations. For instance, the attacker (with malicious intent towards the victim) can strategically select and spoof the location of an e-scooter(s) to a secluded area or an area with minimal human presence to entice victim riders. The attacker (with intent  of financial gain) can follow a similar approach to spoof the e-scooters to a randomized location (or physically hide it plain sight) making it difficult for riders (and e-scooter chargers alike) to find them, leading to financial loss for the service provider. Indirectly, this approach eventually leads to a drained and/or probably stolen e-scooter, and increases the bounty for finding that e-scooter, thus making it a profitable venture for the attacker (e-scooter charger). The attacker (rider) may be able to park in restricted areas or no parking zones, thereby causing public nuisance.   

\subsubsection{Countermeasures} 
To prevent location spoofing, the GNSS modules should not solely rely on GPS but use additional sources for location and to detect spoofing \cite{schmidt2016survey,jansen2019,miralles2018android}. 

\subsection{Fuzzing} 
\label{fuzz}
\subsubsection{Observations}
The fuzzing technique is different from the attack in \cref{dos} where the victim (e-scooter or rider) is overwhelmed with continuous bursts of data rendering them incapable of using the e-scooter. This attack gives the attacker the ability to gauge how each service provider ecosystem handles the e-scooters from the responses, from the API and other control systems, obtained after testing with request or command variants with the intent to identify bugs or vulnerabilities (not found through passive eavesdropping). 
\subsubsection{Consequences} 
The attacker (the rider or the outsider) can infer the protocols used, authentication information (e.g. e-scooter password) and service request-responses by the application by passively eavesdropping or sniffing traffic and actively testing the system, and may be able intercept, manipulate or replay the requests. While the attacker can identify (publicly available) nodes leaking sensitive information, the attacker can also check if the application may be using extraneous permissions that can covertly be exploited to collect information.

\subsubsection{Countermeasures} 
The fuzzing attacks can be controlled by preventing eavesdropping and MITM attacks (\cref{api,mitm}). 

\subsection{User Data Sharing and Inference} 
\label{privacy}
\subsubsection{Observations} 
User data in this micromobility platform can range from any smartphone related activity to a history of locations the rider has visited for any period of time. The micromobility services transfer the data gathered by the applications on the user smartphone over the Internet. \cref{tab:policy} lists the user data collected by these applications as mentioned in their privacy policies. While some applications do not respond to \textit{Do Not Track} requests \cite{razor2019}, most usually share aggregate data with third parties such as Facebook, Google and require location access (Always setting) from users during usage and in background, and therefore the service providers have access to sensitive information (user whereabouts): frequent start and destination places of the riders, a history of locations visited by rider, and possibly their abode. Other data categories and issues are yet to be explored, as they vary from application to application, however the threat that these micromobility services could share private data of users (riders and non-riders) to multiple stakeholders still looms, as which data and what format (raw or processed) these service providers provide them is unknown.

\subsubsection{Consequences} 
Unregulated and non-anonymized data sharing can be used to create a user profile that can later compromise user safety \cite{hallgren2017privatepool,uberBreach-online}. The entity can mine the data obtained from either the service provider or through eavesdropping to infer more information about the micromobility service users and neighboring people. For instance, the attacker (application with intent of financial gain) can discover neighboring users in vicinity via BLE, users in the house via WiFi devices, and use other phone activity usage information to know more about the rider (and user) and his surroundings. The attacker (other stakeholders) can use the information to learn about the users and then strategically place e-scooters on road, entice riders with suitable social media advertisements, etc. The attacker, an outsider with malicious intent, may also be able to gauge the victim's schedule with the location history and be able to identify rider preferences, frequency of visits, and use the information for personal vendetta and financial gain, spreading the information to other malicious entities, who could track the user down in person effectively be a physical threat to the users. 

\subsubsection{Countermeasures} 
Privacy compliance defines how a company conforms to privacy laws, policies, guidelines in different countries or regions when managing confidential personal data \cite{itgovernance_2019}. An extensive analysis, with existing approaches and techniques such as inferring traffic flows of the applications \cite{zhao2019geo,gandhi2018}, is required to check if micromobility services have any information leakages and adhere to what they promise to collect or do with the collected sensitive user data. %Privacy preserving ride sharing \cite{goel2016, nabil2019efficient,hallgren2017privatepool}.

\section{Conclusion}

%loss of money, privacy, physical safety, etc,.
%Even though vandalism brake wires and damage hull. \cite{brakeCut:online}
After describing the various components in the micromobility e-scooter ecosystem, we systematically summarized critical security and privacy concerns in the ecosystem, which both the micromobility users and service providers should be aware of. This first-of-its-kind systematized discussion shall be helpful to developers and researchers for identifying new weaknesses, and improving overall security and privacy properties of the micromobility e-scooter ecosystem. Given the constant evolution and changes in the applications and services, we intend to extensively analyze the ecosystem again in the future.

\balance
\bibliographystyle{acm}
\bibliography{scooters}

\end{document}